\documentclass[9pt,twocolumn,twoside]{pnas-new}

\templatetype{pnasresearcharticle} 

\newcommand{\beginsupplement}{%
        \setcounter{table}{0}
        \renewcommand{\thetable}{S\arabic{table}}%
        \setcounter{figure}{0}
        \renewcommand{\thefigure}{S\arabic{figure}}%
        \setcounter{equation}{0}
        \renewcommand{\theequation}{S\arabic{equation}}%
     }

\title{Functional modules from variable genes: Leveraging percolation to analyze noisy, high-dimensional data}

\author[a,b]{Steffen Werner}
\author[a,1]{W Mathijs Rozemuller}
\author[c,1]{Annabel Ebbing}
\author[c,d,1]{Anna Alemany}
\author[a]{Joleen Traets}
\author[a]{Jeroen S. van Zon}
\author[c,d]{Alexander van Oudenaarden}
\author[c]{Hendrik C. Korswagen}
\author[b,e]{Greg J. Stephens}
\author[a,2]{Thomas S. Shimizu}

\affil[a]{AMOLF Institute, Science Park 104, 1098 XG Amsterdam, The Netherlands}
\affil[b]{Department of Physics and Astronomy, Vrije Universiteit, De Boelelaan 1081, 1081 HV Amsterdam, The Netherlands}
\affil[c]{Hubrecht Institute-KNAW (Royal Netherlands Academy of Arts and Sciences) and University Medical Center Utrecht, Uppsalalaan 8, 3584 CT Utrecht, The Netherlands}
\affil[d]{Oncode Institute}
\affil[e]{Okinawa Institute of Science and Technology, 1919-1 Tancha, Onna-son, Okinawa 904-0495, Japan}

\leadauthor{Werner} 

\significancestatement{
Gene expression largely determines the fate of each cell and ultimately the development and behavior of the whole organism. Whereas most of our knowledge on gene regulatory networks has been obtained from perturbation experiments (e.g.~manipulating environmental conditions, genotype, or other physiological variables), here we develop an alternative approach based on the analysis of naturally occurring variations across individuals within a population. Using both single-cell and whole-animal RNA sequencing data, we demonstrate how a rich set of co-regulated gene modules can be uncovered from transcriptomic variability of individuals within unperturbed populations. To robustly extract interpretable clusters from the strong noise background, we devise a novel, versatile clustering approach based on network theory. With a foundation in the generic behavior of random networks near their percolation critical point, our method is broadly applicable, beyond gene expression, to any noisy, high-dimensional data that sample variation across individuals within a population. 
}

\authorcontributions{G.J.S. and T.S.S. conceived and designed the project. S.W. developed and performed the analysis. W.M.R., A.E. and A.A. conducted the experiments. All authors discussed and interpreted results. S.W., G.J.S. and T.S.S. wrote the manuscript.}
\authordeclaration{The authors declare no competing interest.}
\equalauthors{\textsuperscript{1} W.M.R., A.E. and A.A. contributed equally to this work.}
\correspondingauthor{\textsuperscript{2}To whom correspondence should be addressed. E-mail: shimizu@amolf.nl}


\begin{abstract}
While measurement advances now allow extensive surveys of gene activity (large numbers of genes across many samples), interpretation of these data is often confounded by noise --- expression counts can differ strongly across samples due to variation of both biological and experimental origin. Complimentary to perturbation approaches, we extract functionally related groups of genes by analyzing the standing variation within a sampled population. To distinguish biologically meaningful patterns from uninterpretable noise, we focus on correlated variation and develop a novel density-based clustering approach that takes advantage of a percolation transition generically arising in random, uncorrelated data. We apply our approach to two contrasting RNA sequencing data sets that sample individual variation --- across single cells of fission yeast  and whole animals of {\it C. elegans} worms --- and demonstrate robust applicability and versatility in revealing correlated gene clusters of diverse biological origin, including cell cycle phase, development/reproduction, tissue-specific functions, and feeding history. Our technique exploits generic features of noisy high-dimensional data and is applicable, beyond gene expression, to feature-rich data that sample population-level variability in the presence of noise. 
\end{abstract}

\dates{This manuscript was compiled on \today}
\doi{\url{www.pnas.org/cgi/doi/10.1073/pnas.XXXXXXXXXX}}

\begin{document}

\maketitle
\thispagestyle{firststyle}
\ifthenelse{\boolean{shortarticle}}{\ifthenelse{\boolean{singlecolumn}}{\abscontentformatted}{\abscontent}}{}

A cornerstone of experimental biology is the perturbation-response paradigm, in which targeted manipulations are carefully designed to yield functional and mechanistic insights. With the recent advent of high-throughput techniques, however, the analysis of naturally occurring patterns of variation is emerging as a powerful complementary approach, and has been successfully applied to a variety of problems including protein structure-function mappings \cite{halabi2009protein}, gene-network prediction \cite{munsky2012using}, transgenerational memory \cite{perez2017maternal}, and aging \cite{kirkwood2005accounts}. 

For studies of gene regulatory interactions, a key high-throughput technology is RNA sequencing (RNAseq), which allows transcription-level profiling of gene expression on a genome-wide scale. RNAseq experiments conforming to the perturbation-response paradigm --- differential analysis of gene expression between manipulated and control conditions --- have already transformed our understanding of a wide range of biological processes \cite{ideker2001new, han2015advanced, anders2013count}. With advances in single-cell techniques, RNAseq studies increasingly exploit, beyond perturbation-response, information carried by natural variation across individuals within unperturbed populations. A major success has been in classifying cells within a heterogeneous population into distinct cell types according to transcriptomic differences \cite{grun2015design, kulkarni2019beyond, luecken2019current, liu2016single, wagner2016revealing, andrews2018identifying}.

In this study, we address the complementary challenge of identifying the underlying regulatory relationships among genes from the standing variation in expression across sampled individuals.  Rather than seeking to fully infer the underlying gene regulatory network topology from this inherently (and often prohibitively) noisy class of data \cite{liu2016single, chen2018evaluating}, we focus on identifying functional modules --- sets of genes that demonstrate significant evidence for co-regulation. Extracting gene modules from standing variation can be addressed by clustering expression patterns across samples, and has been attempted in the past with varying degrees of success \cite{kotliar2019identifying, xue2013genetic, shalek2014single, bhosale2013predicting, chen2015spatially, saint2019single, andrews2018identifying}. Yet a primary challenge remains to distinguish true regulatory relationships from noise, and these efforts have depended on expert insights about the specific biological systems to appropriately pre-filter genes, tune analysis parameters, and filter results.

We have developed a novel, data-driven approach motivated by the theory of percolation on random graphs \cite{penrose1995single, penrose2003random, dall2002random, newman2018networks}. The method is conceptually simple yet robustly applicable, reliably yielding interpretable gene clusters across diverse data sets without fine-tuned optimization and filtering steps.
We exploit the generic behavior of random geometric networks close to the percolation critical point, from which we devise a null model for the noise. This noise model in turn provides a basis for identifying statistically significant branches within the cluster hierarchy. 

We demonstrate the robust utility of our approach using two contrasting data sets: one representing single-cell variability across populations of fission yeast (data from ref.~\cite{saint2019single}) and another that samples whole-animal variability across populations of {\it C. elegans} worms (data newly acquired in this study). The yeast single-cell data represent a particularly challenging example for gene clustering due to low mRNA yields from these cells, yet our method successfully extracts multiple functional modules. The results with {\it C. elegans} extend the paradigm of leveraging standing variation to the whole-animal level, yielding at a negligible false discovery rate gene modules of strong functional coherence (\textit{i.e.} permitting internally consistent interpretations across available data sets and annotations) with precise mappings to specific body regions.

\begin{figure*}[htbp]
\includegraphics[width=1\linewidth]{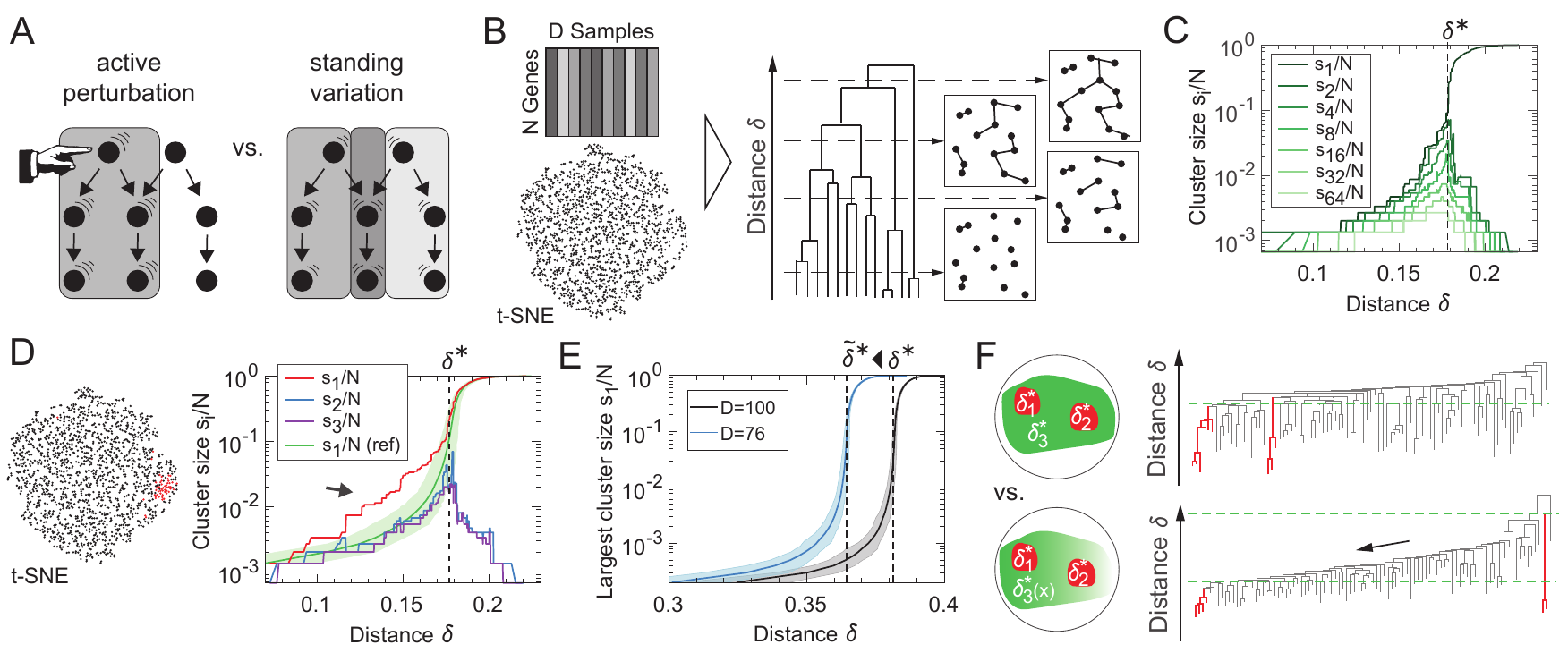}
\caption{{\bf Exploiting standing variation in gene expression by a clustering approach based on network theory.} (A) Illustration of the difference between the canonical perturbation-response experimental paradigm (left), and our approach that leverages standing variation in unperturbed samples (right), which is able to discriminate several co-regulated modules (gray shadings) in the network of genes (dots).
(B) Genes are characterized by a row-vector in the $N\times D$ expression-count matrix. By defining a correlation-based distance measure $\delta$, we associate the degree of co-variation between genes with a distance between respective points in a high-dimensional space.
As an example, we show a t-SNE projection of $N=1500$ genes (points) with a random (normally distributed) expression across $D=10$ measurements.
Cluster formation is traced across the complete single-linkage hierarchical dendrogram. The simplified scheme illustrates how the merging of branches in the dendrogram at distances $\delta$ corresponds to the linking of points (genes) to form clusters.
(C) The largest cluster sizes (here up to rank 64) in uncorrelated data show a generic percolation transition (dashed line) at $\delta^*$,
beyond which the largest cluster size rapidly increases, while all other cluster sizes decrease (same data as in t-SNE of (B)).
(D) We replace 50 genes in the uncorrelated data of (B) by correlated genes (red, see SI). The growth of the corresponding cluster (arrow) is significantly faster than that expected for the largest cluster size of completely uncorrelated data (green, shading denotes 3 standard deviations in $\delta$).
(E) Restriction of the data to a lower dimensional space (blue) shifts the percolation transition from $\delta^*$ to a smaller value $\tilde{\delta}^*$.
(shading marks 3 standard deviations in $\delta$, $N=10^4$).
(F) As the noise in gene expression data is typically inhomogeneous, our clustering probes the local percolation behavior (bottom) instead of a global percolation transition (top) to distinguish clusters of correlated genes (red) from spurious clusters.}\label{fig:method}
\end{figure*}

\section*{Results}
\subsection*{Percolation approach to clustering noisy, high-dimensional data}
We leverage the standing variation across unperturbed samples to reveal functional modules in gene regulatory networks (Fig.~\ref{fig:method}A).
Groups of functionally related genes are expected to share a common pattern of expression variation across samples, the similarity of which can be quantified by a correlation-based distance measure $\delta$ (Fig.~\ref{fig:randomnetworks}A), ranging from $0$ (perfect correlation) via $1/2$ (no correlation) to $1$ (perfect anti-correlation). 
Gene expression data are provided in the form of a $N\times D$ expression-count matrix for $N$ genes measured across $D$ samples (Fig.~\ref{fig:method}B, top left).  Each gene can then be considered a point in the $D$-dimensional space of possible expression profiles, with the coordinate along each of $D$ dimensions given by the corresponding row within the expression-count matrix.
In Fig.~\ref{fig:method}B (bottom left) we use a 2-dimensional t-SNE projection to visualize the distribution of such points for a synthetic data set of $N=1500$ uncorrelated genes with random (normally distributed) expression values across $D=10$ measurements (see SI).

We use single-linkage hierarchical clustering as a basis to explore the data by keeping track of all clustering events as $\delta$ is increased, from the first grouping of two genes to the final merging of all data points into a single giant cluster (Fig.~\ref{fig:method}B, right).
The graph resulting from this procedure for uncorrelated normally distributed genes yields a geometric random network (Fig.~\ref{fig:randomnetworks}), which exhibits a percolation transition with generic behavior as the linkage threshold distance $\delta$ is varied \cite{dall2002random, hartigan1981consistency, penrose1995single, penrose2003random}.
The transition is readily observed by ranking (in decreasing order) the sizes $s_i$ of all clusters obtained for each value of $\delta$, and tracking these ranked cluster sizes as a function of $\delta$ (Fig.~\ref{fig:method}C). Initially, all cluster sizes tend to increase, however at a critical value $\delta^*$, we observe the percolation transition, beyond which only the largest cluster size $s_1(\delta)$ rapidly grows at the cost of all the other clusters. Importantly, the critical distance $\delta^*$ as well as the manner in which cluster sizes grow as $\delta$ approaches $\delta^*$ is a generic property of the percolation transition, and is completely specified by $N$ and $D$ for uncorrelated and normally distributed data (Fig.~\ref{fig:randomnetworks}). 
The cluster growth as a function of $\delta$ changes if the data contain correlated genes, which form groups of points of increased density as exemplified in Fig.~\ref{fig:method}D (left, red points). The linkage of points (\textit{i.e.}~genes) within such a group is governed by a smaller percolation threshold in comparison to $\delta^*$ of the noise background. Thus, for sufficiently strong correlations, the size of a cluster containing those correlated genes significantly exceeds the expected largest cluster size of normally distributed genes at some value of $\delta<\delta^*$ (Fig.~\ref{fig:method}D right, arrow). By analyzing the cluster growth as a function of $\delta$, clusters of correlated genes can be discriminated from spurious clusters arising from linkage of the noise background.

The assumption of uncorrelated and normally distributed noise is not accurate for real-world gene expression data. Stochastic gene expression levels do not typically follow a normal distribution, and experimental protocols such as RNA count extraction as well as data pre-processing steps (\textit{e.g.}~filtering and normalization) can additionally introduce non-trivial interdependencies of between some or all of the genes in the data set \cite{grun2014validation, hafemeister2019normalization, vallejos2017normalizing}. Consistent with such deviations from the assumed noise model, in typical RNAseq data we observe a percolation transition at values of $\delta^*$ smaller than expected for uncorrelated and normally distributed noise, given the data set size ($N\times D$). 

We approximate the noise background by uncorrelated and normally distributed noise of reduced dimensionality $\tilde{D}<D$, which shifts the percolation transition to a smaller correlation distance $\tilde{\delta}^*<\delta^*$ due to the (on average) decreased distance between points. By tuning $\tilde{D}$, we can thus calibrate the uncorrelated and normally distributed noise model to match its critical point to that of the measured data (\textit{i.e.}~$\delta^*\rightarrow\tilde{\delta}^*$). In the SI, we show that this calibration by $\tilde{D}$ provides a conservative estimate for the largest cluster size expected from noise (Fig.~\ref{fig:randomnetworks}M), and that the resulting percolation behavior is often nearly indistinguishable from other potentially more realistic choices for the null model (Fig.~\ref{fig:randomnetworks}N).

An additional complication in real-world data is that correlations of the background noise (modeled here as an effective decrease in dimensionality) typically vary across the data set. When such variations in the global correlation level are negligible (Fig.~\ref{fig:method}F, top) a single, global percolation threshold suffices to identify significant clusters. Otherwise, the global percolation transition is smeared out \cite{hebert2019smeared} and a single constant value for $\delta^*$ does not faithfully characterize the noise background. To account for such inhomogeneities, we apply our percolation analysis iteratively across the dendrogram (from high to low $\delta$ values) at each branch point (see SI). This iterative scheme enables detection of significant clusters even when their respective branchpoints are separated by large differences in the correlation distance $\delta$ (Fig.~\ref{fig:method}F, bottom).

\begin{figure}[tb]
\includegraphics[width=1\linewidth]{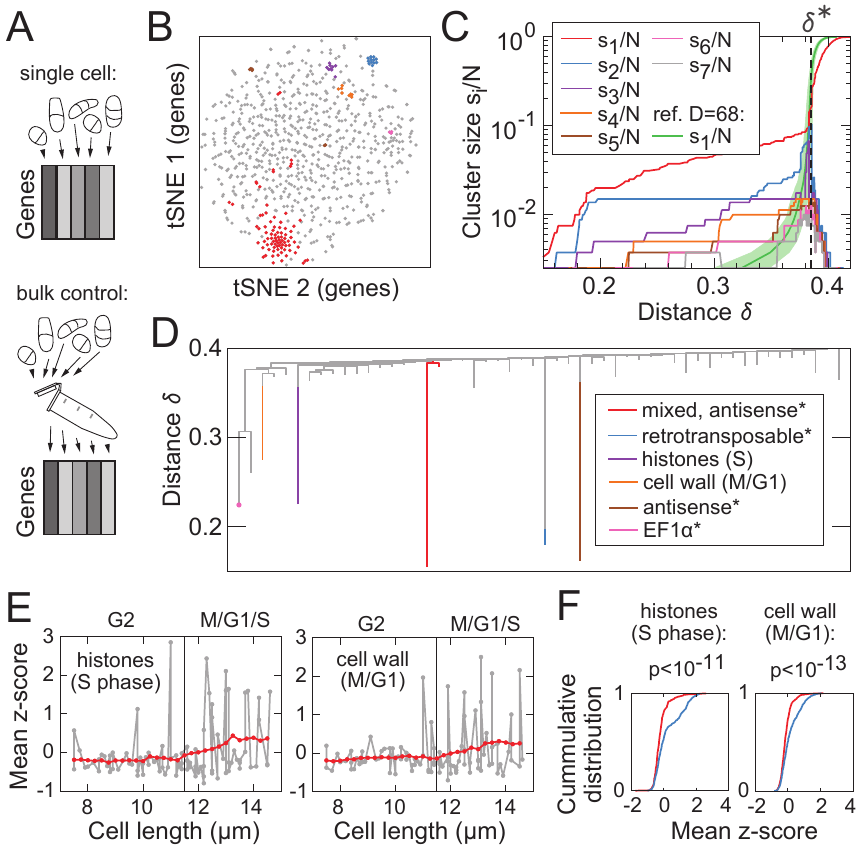}
\caption{{\bf Clustering single-cell expression variation detects functional modules despite a noisy background.} (A) We analyze 9 data sets (with $D=96$) from ref.~\cite{saint2019single} with individual fission yeast cells and 7 control data sets of bulk sequencing data (with $D=96$). (B) The t-SNE projection of the genes in data set 1 (with perplexity 35) illustrates the six subsequently found clusters (colors) hidden in a very noisy background (gray). (C) The global percolation threshold $\delta^*$ in data set 1 compares to the percolation threshold in lower dimensional uncorrelated data with $D=68$ (green, shading denotes 3$\sigma$ interval).
Six colored curves significantly exceed the expected largest cluster size, suggesting that six clusters of correlated genes can be found.
(D) Dendogram with identified clusters (same color code as in (B)) and their prevalent gene annotation (asterisks denote overlap with clusters in control data set). 
We use a significance threshold of 3$\sigma$ in $\delta$ for our clustering method probing the local percolation behavior.
(E) Gene expression dynamics during cell cycle progression for the two clusters exclusively found in single cell data sets (red: moving averages in length bins of $1\mu m$).
(F) The distribution of expression values in G2 (red) is significantly different from the distribution in the other cell cycle stages (blue) when pooling data from all single cell data sets, assessed by a Kolmogorov-Smirnov test.}\label{fig:yeast}
\end{figure}

\subsection*{Clustering single-cell expression variation in fission yeast}
As a first example, we apply our clustering method to RNAseq data of the fission yeast {\it Schizosaccharomyces pombe} from Saint {\it et al.}~\cite{saint2019single}. Because achievable mRNA yields from individual yeast cells are inherently low, these data exhibit a particularly high level of counting noise and represent a challenging case for cluster identification (analogous to that illustrated with synthetic data in Fig.~\ref{fig:method}D). The yeast cells are not synchronized and we thus anticipated that we might identify gene clusters associated with cell cycle progression. For this analysis, we chose from ref.~\cite{saint2019single} 9 single-cell data sets obtained under 6 different conditions, each of which comprises $D=96$ cells, as well as 7 control data sets where bulk sequencing was performed on cell populations (Fig.~\ref{fig:yeast}A). The t-SNE projection in Fig.~\ref{fig:yeastall}A shows that cells from the same data set cluster together. Thus, we analyze each of the 16 cell and control data sets separately, followed by a cross-validation of the individual results. As a representative example, we focus here on data set 1, and provide results across all data sets in Fig.~\ref{fig:yeastall}.

The t-SNE plot of the genes in Fig.~\ref{fig:yeast}B illustrates the noisy data structure analogous to Fig.~\ref{fig:method}B with the six subsequently identified clusters (colored points) immersed within a dominant noise background (gray points).
Clustering of these points reveal a percolation transition in Fig.~\ref{fig:yeast}C at a value $\delta^*$ much smaller than that expected for uncorrelated data. This transition threshold is well matched by that of an uncorrelated and normally distributed noise model of reduced dimensionality $\tilde{D}=68$ (Fig.~\ref{fig:yeast}C, green curve) compared to that of the data ($D=96$). Each of the colored curves indicating the evolution of top-ranking cluster sizes $s_1-s_6$ in Fig.~\ref{fig:yeast}C are clearly above the expected largest cluster size from uncorrelated data (green curve), indicating the existence of six significant clusters.

Indeed, by applying our clustering technique, we obtain six clusters (colored branches in the dendrogram of Fig.~\ref{fig:yeast}D), which on the basis of member gene annotations appear functionally distinct from one another and are found to recur across several data sets  (Fig.~\ref{fig:yeastall}G). Yet the results also illustrate the challenge of extracting the extremely faint signals in yeast single-cell data. Four of the six obtained clusters  (asterisked in Fig.~\ref{fig:yeast}D) are associated with non-coding RNA and/or are also found in the bulk control data sets, suggesting that these clusters likely reflect experimental, rather than biological variation. For example, the sequencing counts of several clusters depends on the specific well of the 96-well plate and correlates with the total extracted RNA count (Fig.~\ref{fig:yeastall}H). The two clusters in Fig.~\ref{fig:yeast}D exclusively found in single-cell data are, according to member gene annotations, associated with specific cell cycle stages. One cluster almost exclusively consists of 8 histone genes (all 9 histone genes of yeast are found, when combining the analysis of all data sets) and is thus up-regulated in S phase. Interestingly, for the histone pair H3.2/H4.2 (which encodes the same amino acid sequence), a relatively constant expression level was previously reported \cite{takayama2007differential}, yet we find co-variation of these genes with other histones that are known to peak in S phase (Fig.~\ref{fig:yeast}E, left). The second cluster exclusive to single-cell data is associated with modifications in rigidity of the cell wall, especially in M and G1 phase \cite{takada2010cell, popolo2001yeast,duenas2010characterization, lock2018pombase,jaiseng2012studies}, and also shows the expected cell-cycle dependent expression (Fig.~\ref{fig:yeast}E, right). 
We can further confirm the cell-cycle dependent cluster behavior by combining the results from all data sets. We merge the respective clusters found independently across different data sets (Fig.~\ref{fig:yeast}G) and compare the expression distributions of the G2 phase (red) and the other cell cycle stages (blue) in Fig.~\ref{fig:yeast}F.
Taken together, these results demonstrate that despite the strong noise background in single-cell RNAseq data, our analysis approach can robustly identify multiple meaningful gene clusters from the standing variation in an unperturbed cell population.

\begin{figure}[tb]
\includegraphics[width=1\linewidth]{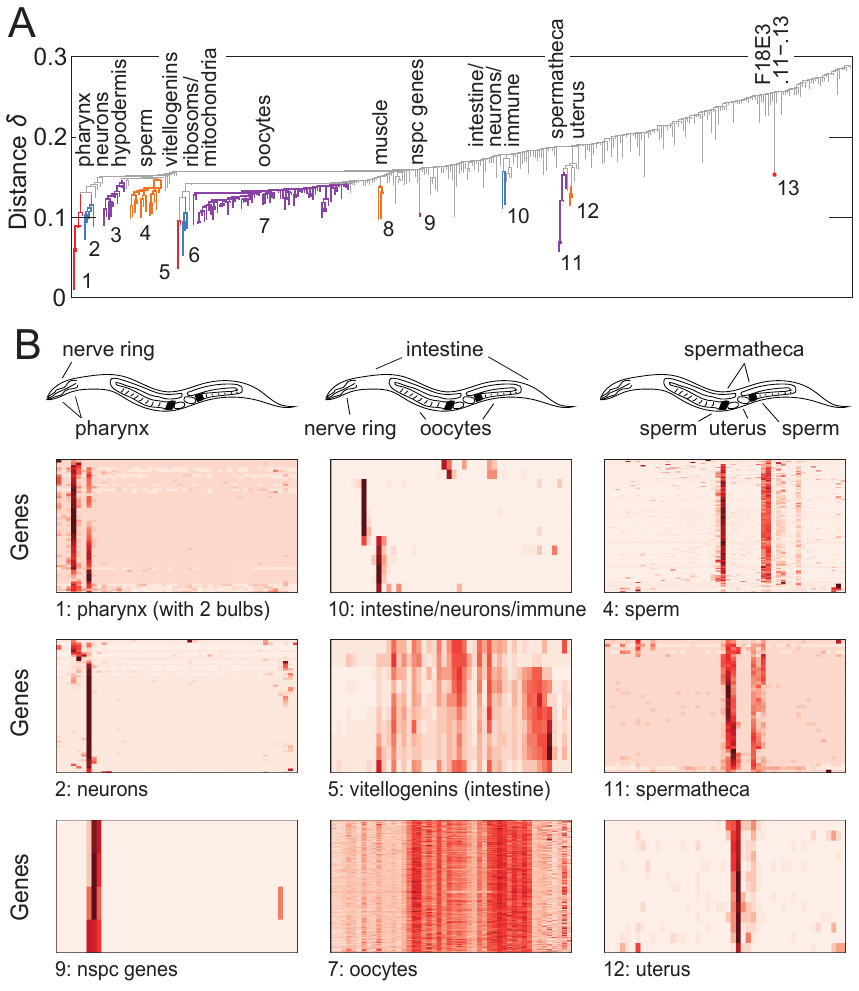}
\caption{{\bf Clustering whole-animal expression variation reveals tissue-specific gene modules.} (A) Our clustering approach of whole-worm RNAseq data from young adult hermaphrodites (with $D=34$) identifies 13 clusters (colored branches) with tissue-specific annotation. (B) Tomoseq data of worm slices from head to tail of ref.~\cite{ebbing2018spatial} reveal distinct spatial patterns of our identified clusters from (A) associated with specific anatomical regions. We display the expression of all genes of the non-homogenous clusters, where red denotes high expression and white low expression values.}\label{fig:celegans}
\end{figure}

\subsection*{Clustering whole-animal expression variation in \textit{C. elegans}}
Having determined that functional gene modules can be identified from expression variations across populations of single cells, we ask whether the same approach could be fruitfully extended to the whole-organism level.
To answer this question, we perform RNAseq on whole animals of the nematode \textit{C. elegans}, to generate a data set comprising $D=34$ individual (genetically identical) worms in the young adult stage. In contrast to the non-synchronized yeast data, these worms are synchronized at hatching and grown under identically controlled conditions throughout the course of development (see Methods). We identify 13 gene clusters associated with specific functions and distinct anatomical regions (Fig.~\ref{fig:celegans}A). A tissue-specific annotation is assigned to each of these clusters by mining the expansive literature on {\it C. elegans}, with particularly extensive use of the compendium of annotated gene data on WormBase \cite{angeles2016tissue, angeles2018two} as well as a recently published spatially resolved RNA tomography (tomoseq) data set by Ebbing \textit{et al.}~\cite{ebbing2018spatial}.
The tomoseq data provide spatial maps of gene expression in individual slices along the worm. In (Fig.~\ref{fig:celegans}B) we illustrate the tissue-specific expression patterns for 9 inhomogeneously expressed gene clusters (see Fig.~\ref{fig:tomoseq} for a comparison of all 13 clusters across four different individuals).
For example, the genes of cluster 1 are expressed almost exclusively in the pharynx with the two pharynx bulbs clearly visible, the genes of cluster 2 are located in the brain, and the genes of cluster 4 are found in two stripes flanking the spermatheca (cluster 11) where sperm is produced. Additionally, the tissue, phenotype and GO-term enrichment analysis on WormBase allows for cross-validation and further characterization of the clusters. For example, 83 of the 108 genes of the globally expressed hypodermis cluster 3 are annotated as epithelial genes and 790 out of 794 genes in the oocytosis cluster 7 are associated with reproduction \cite{angeles2016tissue, angeles2018two}. Collectively, these results demonstrate a high degree of functional coherence among genes within essentially all identified clusters, thus indicating that the effective false discovery rate for both cluster significance and cluster membership are remarkably low.

These gene modules identified by our approach and cross-validated against prior data reveal new insights into the genetic network of {\it C. elegans}. For example, cluster 5 contains all six known vitellogenins but in addition also four genes of unknown function. The joint cluster membership, as well as the tomoseq expression patterns, indicate that these scarcely characterized genes are expressed together with vitellogenins in the intestine and thus likely involved in yolk formation. Similarly, only very few of the 39 genes in cluster 11 are known to be specific to the spermatheca (see SI Data Table 3), however our clustering result in conjunction with the validation by tomoseq data strongly indicate that the remaining co-varying and co-localized genes are also associated with the spermatheca. Even for well characterized genes, we find new and unexpected functional associations. The pharynx cluster 1 not only contains many pharynx-specific muscle genes (and the marginal cell marker \textit{marg-1}) but also several genes related to stress and immune responses. Cluster 6 contains genes for both a large number of ribosomal proteins as well as proteins localized to the mitochondria. Finally, the genes in cluster 10 are present in the most anterior intestinal cells as well as some neurons, and thus might together be responsible for an innate immune response \cite{pukkila2012stimulation, angeles2016tissue, angeles2018two}.

\begin{figure}[tb]
\includegraphics[width=1\linewidth]{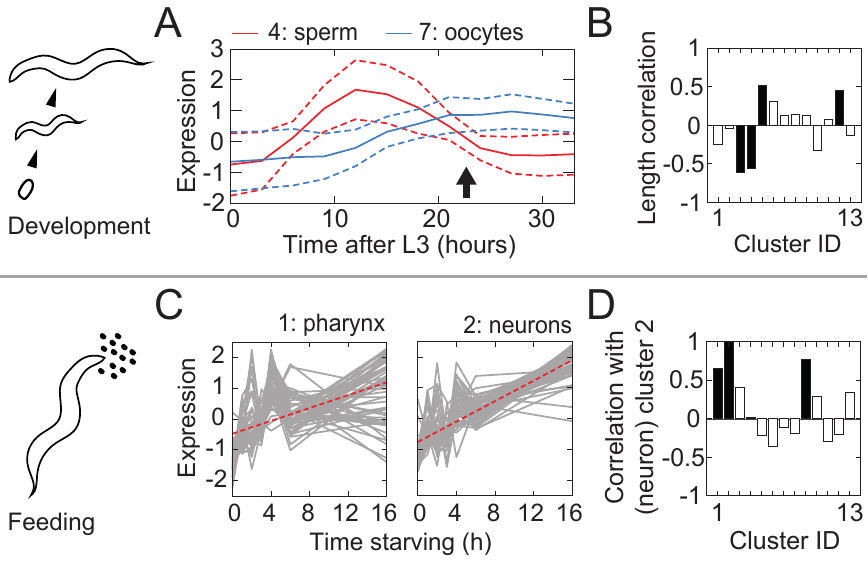}
\caption{{\bf Clusters identified from standing variation represent gene modules co-regulated in different functional contexts.} (A) Developmental dynamics of our identified spermatogenesis and oogenesis clusters based on data from ref.~\cite{reinke2004genome} (solid line: mean, dashed lines: standard deviation, arrow marks approximate time point of our measurement). (B) Four of our clusters correlate significantly (p<0.01) with length of the worm (black). (C) The two clusters of neuron and pharnyx genes increase in response to starvation, analysis of data from ref.~\cite{harvald2017multi}.
(D) Besides cluster 9 (consisting of only 4 nspc genes), the pharynx cluster 1 is the only one significantly correlated with the neuron cluster 2 in our measurement (black: p<0.01).}\label{fig:transgen}
\end{figure}

The successful identification of these functional gene modules demonstrates the power of analyzing standing variation at the whole-organism level, but leaves open an important question: what are the origins of these informative variation patterns, which manifest under identically controlled laboratory conditions? One potential source of variability is differences in worm development.
If the expression of a set of genes rapidly changes at the young adult stage where the measurement takes place, small differences in worm development (which exist at a finite level even in synchronized populations) could translate into significant variations in gene expression across individuals. As an example, in Fig.~\ref{fig:transgen}A we utilize developmental transciptomic data by Reinke {\it et al.} \cite{reinke2004genome} to visualize the temporal dynamics of the two germ line related clusters, cluster 4 (spermatogenesis) and cluster 7 (oogenesis) identified in our own data. The arrow marks the approximate time point of our measurement, at which sperm development is near completion but still changing, while the oogenesis genes have already reached a plateau. This suggests that the variations of the spermatogenesis cluster might at least partially arise from a developmental effect, which however cannot explain the emergence of the oogenesis cluster. In Fig.~\ref{fig:transgen}B we show the correlation between gene expression and worm length as a proxy for the developmental progress in our data set \cite{gritti2016long}. Indeed, the spermatogenesis cluster decreases with worm length, yet the oogenesis cluster is not significantly correlated. Further clusters with strong length correlations are cluster 3 (hypodermis), which is known to oscillate during development and is expected to be downregulated in the measurement window, and cluster 5, which is associated with yolk provisioning in the eggs and known to vary with maternal age \cite{perez2017maternal, jordan2019insulin}. Finally, the observation that the uterus genes (cluster 12), yet not the spermatheca genes (cluster 11) are correlated with worm length is in agreement with fluorescence data from ref.~\cite{zimmerman2015reproductive}, showing that ule-3/ule-5 from cluster 11 is not visible before adulthood in contrast to ule-2/ule-4 from cluster 12.

A second potential source of expression variability is the individual's experience of external stimuli. Feeding conditions, in particular, are important drivers of {\it C. elegans} behavior. Our analysis of RNAseq data from Harvald \textit{et al.}~\cite{harvald2017multi} demonstrate that the pharynx genes in cluster 1 and the neuron genes in cluster 2, as well as the small cluster 9 (also expressed in neurons) respond to an extended starvation period in an analogous manner (Fig.~\ref{fig:transgen}C and Fig.\ref{fig:celeganspre}E). Interestingly, these clusters are on average also strongly correlated with each other within our data set (despite being found independently), Fig.~\ref{fig:transgen}D. 
As our experiments were designed to sample standing variation, worms were not specifically manipulated to test for feeding-dependent effects and the recorded small variations in the $\sim 1$~hr starvation period before sequencing does not correlate significantly with expression variations (see SI). However it remains a possibility that cumulative differences in each worm's feeding behavior from further in the past contributes to the observed expression variations. Consistent with this view, we find a large overlap of more than 60\% of the neuron genes from cluster 2 with a gene set from Freytag {\it et al.} implicated in starvation memory \cite{freytag2017genome} (Fig.~\ref{fig:celeganspre}F). Further, we find hints of a transgenerational memory effect from RNAseq measurements in L1 worms by Webster {\it et al.}~\cite{webster2018transgenerational}. Genes of the neuron-associated cluster 2 (and possibly also those of the pharynx-related cluster 1) are up-regulated in worms whose ancestors (three generations ago) were subject to severe starvation conditions in the experiments of ref.~\cite{webster2018transgenerational} (Fig.~\ref{fig:celeganspre}G). Altogether, this series of observations on cluster 1, 2, and 9 suggest that their correlated expression variation might encode differences in integrated feeding experience (potentially even across generations).

Overall, our RNAseq experiments and analysis in \textit{C. elegans} demonstrate how various functionally interpretable gene modules can be extracted from standing variation at the whole-animal level.

\section*{Discussion}
By developing a novel clustering approach based on the network theoretical idea of random graph percolation \cite{newman2018networks, penrose2003random, dall2002random, penrose1995single}, we have demonstrated that functionally related sets of genes can be robustly extracted from the standing expression-level variation across individuals within unperturbed populations.

A key feature of our approach is that we do not assume that the entire data set can be partitioned into meaningful clusters. Instead, we focus on identifying the subset of gene clusters that are most likely to carry interpretable biological information, discarding the remaining data as noise. Density fluctuations across the data set are accounted for by applying the percolation-based thresholding analysis locally, at every branch point of the cluster hierarchy. In this respect, there are similarities to approaches such as the dynamic tree cut method \cite{langfelder2007defining}, hierarchical density-based clustering \cite{McInnes2017}, the search for density-peaks \cite{rodriguez2014clustering, marques2018clusterdv} and graph-based bi-clustering \cite{maere2008extracting}.
A unique feature of our method is a solid grounding in the statistical physics of percolation. By exploiting the generic behavior of random networks close to the percolation critical point, it provides an unambiguous framework for significance testing of clusters without the need of arbitrary parameter choices.
A single parameter controls for the false discovery of clusters, for which a standard choice ensures that essentially all obtained clusters tend to be biologically interpretable.

Gene clustering can be especially challenging within single-cell data, due to inherently high noise levels associated with low molecular counts \cite{liu2016single, grun2014validation, kharchenko2014bayesian}. Other methods typically rely on pre-processing steps to deal with this noise background, in particular by filtering the data set to retain only the most strongly varying genes for analysis \cite{kotliar2019identifying,shalek2014single,saint2019single}. Our approach does not exclude \textit{a priori} individual genes on the basis of low variance and is thus able to identify modules consisting even of weakly varying genes, as exemplified in our analysis of the fission yeast data of ref.~\cite{saint2019single}  (Fig.~\ref{fig:yeast}, SI Data Table 1-2). These data provided a particularly stringent test for our gene-clustering approach, given the inherently low mRNA yield, yet our method nevertheless succeeded in extracting functionally interpretable gene clusters from the standing expression-count variation (Fig.~\ref{fig:yeast}). Interestingly, the significant gene clusters we identified as related to cell length (Fig.~\ref{fig:yeast}D-F) are not enriched in the set of genes identified in ref.~\cite{saint2019single} as related to that phenotype. The difference likely reflects the complementarity of approach --- whereas in ref.~\cite{saint2019single} individual genes were directly compared to the cell cycle phase, here we asked whether and how gene modules identified by clustering relate to the same phenotype. 

To our knowledge, extracting gene modules from standing variation has rarely been attempted at the whole-organism level, the whole-plant RNAseq experiments of ref. \cite{bhosale2013predicting} being a notable exception. Our RNAseq experiments on {\it C. elegans} worms demonstrated that analyzing the standing variation across whole animals in a population can reveal a rich set of clusters, the regulatory significance of which we verified through both functional annotations and spatial expression patterns of constituent genes (Fig. \ref{fig:celegans}). The diversity of biological functions to which these  clusters could be mapped demonstrates the power of leveraging standing variation, which can be thought of as a high-dimensional superposition of multiple perturbations. 
Biological drivers of variability potentially include intrinsic stochasticity in gene expression, temporal dynamics (e.g. due to development and/or aging) as well as differences in the history of interactions with the environment by each individual \cite{padovan2013using, perez2017maternal, wagner2016revealing, raj2008nature, eldar2010functional}. These diverse effects can be strongly intertwined and integrated over the organism's life history, in some cases even across generations. Analysis of the clusters identified in our \textit{C. elegans} data set demonstrated evidence of such variation in development and environmental interactions (Figs.~\ref{fig:transgen},\ref{fig:celeganspre}).   

Our clustering approach is conservative by design, which means that the effective false discovery rate is low for both cluster identification (\textit{i.e.} essentially no clusters from spurious noise-induced correlations) and cluster membership (\textit{i.e.} very few spuriously added genes within identified clusters). Whereas we only identify gene clusters with correlations strong enough to be distinguishable from the noise background, our experience in applying this approach to diverse RNAseq data sets has been that nearly all identified clusters can be confirmed to reflect true functional gene interactions, either via functional annotations (e.g. GO-term analysis) or through further analysis of the clustered genes in complementary perturbation-response data sets. 
Thus, our approach is particularly well suited for exploratory analyses to uncover co-regulated gene modules within samples with little prior knowledge. The resulting clusters can then serve as a resource for further in-depth investigations. 

Given its foundation in the generic behavior of random networks near their percolation critical point, we expect our method to be broadly applicable beyond the specific context explored here, for example to cluster genes according to spatial, rather than between-individual variation \cite{chen2015spatially,ebbing2018spatial}, and --- beyond gene expression --- to any noisy, high-dimensional data that sample variation across multiple measurements.

\subsection*{Supporting Information (SI)}
\beginsupplement

\begin{figure*}[t!]
\includegraphics[width=0.99\linewidth]{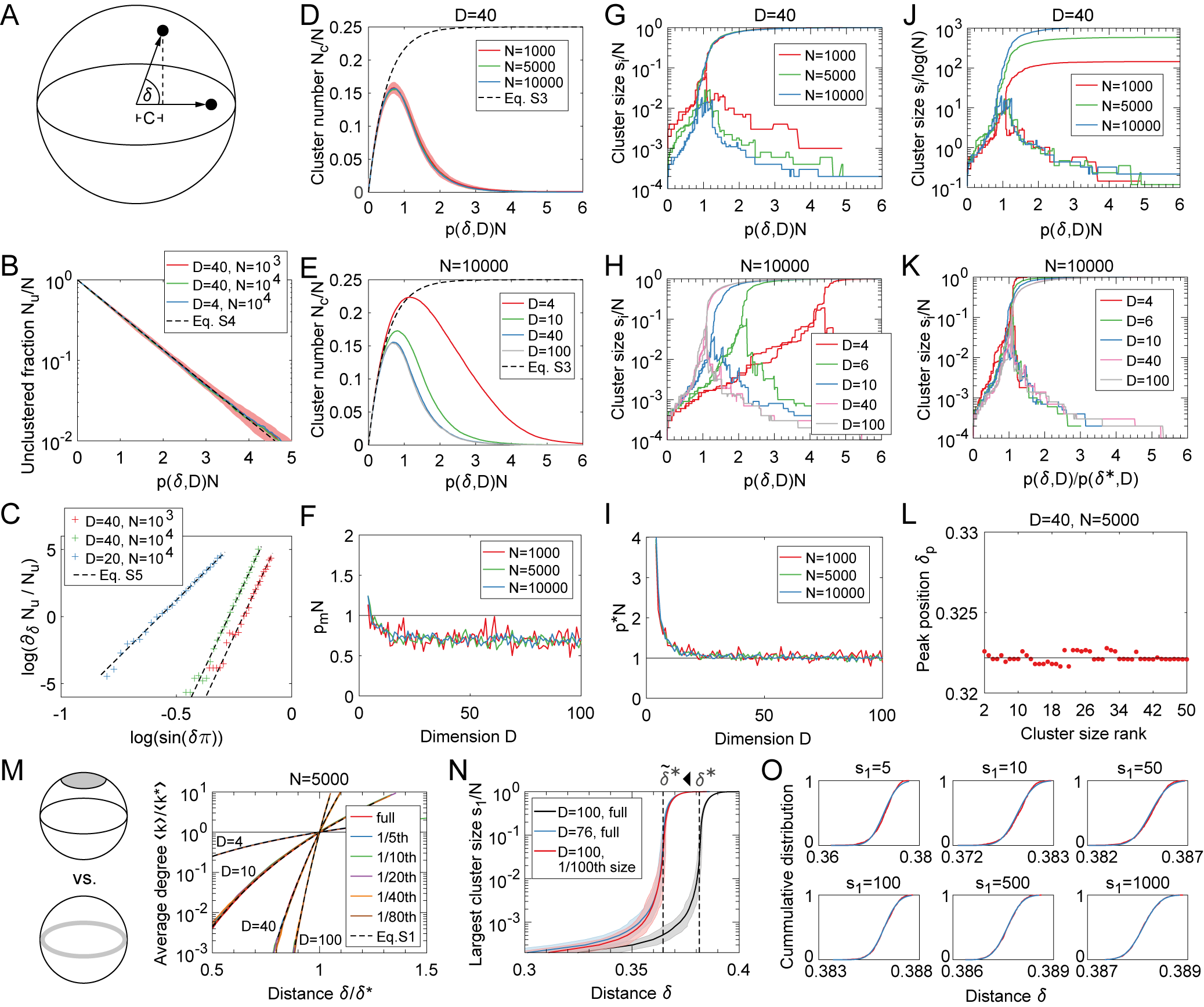}
\caption{{\bf Random geometric networks show generic behavior at the critical point similar to Erd\"os-R\'enyi random networks}
(A) We use a correlation-based distance measure to compare gene expression variation, which allows to represent each gene as a point (black dots) on a $(D-1)$-dimensional hypersphere. We define the angle distance $\delta=\text{acos}(C)/\pi$ based on the Pearson correlation $C$.
(B) Relative number of unclustered nodes (solid lines denote the mean over several realisations, the shading denotes the respective standard deviation for the red curve). (C) Decay rate of unclustered nodes allows to determine $D$ and $N$ from a linear fit. 
(D) Number of clusters as a function of $\langle k\rangle=pN$ shows a generic behavior independently of $N$ (solid lines denote the mean over several realisations, the shading denotes the standard deviation for $N=1000$). (E) Number of clusters as a function of $\langle k\rangle=pN$ approaches a generic behavior for sufficiently large $D$. (solid lines denote the mean over several realisations) (F) Peak position $p_mN=\text{argmax}(N_c)$ of the cluster number as a function of $D$.
(G) The behavior of the two largest cluster sizes $s_1$ and $s_2$ as a function of the mean degree $\langle k\rangle=pN$ shows a generic percolation transition independently of $N$. (H) The behavior of the two largest cluster sizes possesses a percolation transition at $p^*N=1$ for sufficiently large $D$. (I) Critical mean degree as a function of $D$ approaches $1$, obeying a generic behavior $p^*N=f(D)$ independent of $N$.
(J) All cluster sizes before the percolation transition and all but the giant cluster after the percolation transition scale with $log(N)$ (same data as in (G)).
(K) We observe a generic behavior of the cluster sizes independent of $D$, when rescaling by the critical value $p^*$ (or $\langle k \rangle$) (same data as in (H)).
(L) The peak value $\delta_p=\text{argmax}(s_i)$, at which the curve for the largest cluster sizes, for the second largest cluster sizes and for higher ranks of cluster sizes reaches its respective maximum, approaches the percolation threshold $\delta^*$ (solid line).
(M) Rescaling allows to generally describe the relationship between $\delta$ and the average degree $\langle k \rangle$ by the function of Eqs.~\ref{eq:connectionprob}-\ref{eq:meandegree}, even when restricting the nodes to a subspace of original hypersphere, e.g.~a cap (legend specifies surface fraction, curves overlap). The average degree increases with decreasing $D$ for $\delta<\delta^*$.
(N) Example from Fig.~\ref{fig:method}E, showing that not only the restriction of the data to a lower dimensional manifold (blue) but also to a subspace of the original hypersphere (red), for example a cap (spanning only a fraction of the hypersurface) shifts the percolation transition $\delta^*$ to a smaller values $\tilde{\delta}^*$.  The expected largest cluster sizes for all $\delta<\tilde{\delta}^*$ are higher in the lower dimensional manifold in comparison to the cap of matching $\tilde{\delta}^*$, however the differences are small for typical parameter values (shading marks 3 standard deviations, here: $N=10^4$).
(O) The distribution of $\delta$ at which the largest cluster reaches a certain size is well described by a normal distribution. }\label{fig:randomnetworks}
\end{figure*}

\paragraph{Geometric representation of gene expression data:}
The gene expression data are provided in the form of a $N\times D$-matrix of the RNA sequencing counts of $N$ genes in $D$ samples. Each gene can then be considered a point in the $D$-dimensional space of possible expression profiles, defined by a $D$-dimensional row-vector. We compare the similarity in expression variation between genes across samples by a correlation-based measure. By computing the Pearson correlation $C$, we implicitly subtract the mean of each row-vector and scale the vector to unit length. Thus, we remove two degrees of freedom, such that each gene becomes a point lying on the $(D-2)$-dimensional hypersurface of a $(D-1)$-dimensional unit hypersphere (Fig.~\ref{fig:randomnetworks}A).
Points corresponding to highly correlated genes lie in close proximity to one another. The Pearson correlation coefficient $C$ between two gene expression profiles is mathematically equivalent to the scalar projection (dot product) of their respective vectors directed from the center of the hypersphere to their respective points on the hypersurface, and the Pearson similarity distance 1-$C$ widely used in gene expression analysis can be considered a measure of the angle $\theta=\text{acos}(C)$ between these vectors. For the following analysis, the angle itself proves to be a more natural choice for quantifying (dis)similarity, and we define $\delta\equiv\theta/\pi$, which ranges from $0$ (perfect correlation) via $0.5$ (no correlation) to $1$ (perfect anticorrelation) as the distance measure between genes.

We model uncorrelated gene expression data by expression values drawn from a normal distribution, which results in a uniform distribution of the respective points on the hypersphere (irrespective of mean and variance of the normal distribution).

\paragraph{Two types of random networks:}
We perform single-linkage hierarchical clustering by connecting points on the hypersphere as a function of the threshold distance $\delta$. For uniformly distributed data points, the resulting graph is a geometric random network and in the following, we refer to the data points as the nodes of the networks. We can compare the geometric random network to a random network of the well-studied Erd\"os-R\'enyi type,
where each of the $N$ nodes can be connected to any other node with a constant probability $p$ \cite{gilbert1959random,bollobas2001random,newman2018networks, erdos1959random}. This class of random networks without a geometric underpinning contrasts with geometric random networks where connections between pairs of nodes are not independent. For example, two nodes that share a common neighbour in a geometric random network are more likely also close to each other and thus linked \cite{penrose2003random}. However, in the limit of large dimensions $D$, the behavior of geometric random networks becomes equivalent to Erd\"os-R\'enyi networks \cite{dall2002random, devroye2011high, penrose2003random} because, loosely speaking, the connectivity between each pair of nodes tends to be determined by one of the many orthogonal dimensions. For example, the clique number (\textit{i.e.}~size of the largest fully connected cluster) behaves similarly in both random network types for $D>(\log_{10}(N))^3$ \cite{devroye2011high}.

Here, we show that even for small values of $D$, many characteristics of geometric random networks are equivalent to Erd\"os-R\'enyi random networks (at least in the relevant regime of $\delta<\delta^*$) and the probability $p$ that two nodes are connected essentially determines in both network types (i) the number of unclustered nodes, (ii) the number of clusters, and (iii) the percolation behavior.
The pairwise connection probability $p(\delta,D)$ in uniform data is given by the fraction of the hypersurface attributed to a hyperspherical cap of opening angle $\delta$:
\begin{equation}
p=I[\sin(\delta\pi)^2;D/2-1,1/2]/2\quad\text{for }\delta<1/2\,,
\label{eq:connectionprob}
\end{equation}
where $I$ is the incomplete regularized beta function. The average degree (\textit{i.e.}~average number of connections) of a node is given by
\begin{equation}\langle k\rangle=p(N-1)\approx pN\,.
\label{eq:meandegree}\end{equation}

\paragraph{Number of unclustered nodes:}
To first order, the probability that a node is isolated is $P_{iso}=(1-p)^{N-1}$
and the expected number of unclustered nodes is
\begin{equation}
N_u=NP_{iso}=N(1-p)^{N-1}\,.
\end{equation}
In general, we are interested in the limit of small $p$ (at most of the order of $\mathcal{O}(1/N)$) because otherwise (almost) all nodes are part of a cluster. For this limit, we obtain
\begin{equation} \label{eq:NuApprox}
N_u\approx Ne^{-p(N-1)}\,.
\end{equation}
The fraction of unclustered nodes $N_u/N$ only depends on the mean degree $\langle k\rangle$. Interestingly, $D$ can be extracted directly from the exponent of the relative change in $N_u$, which shows a power law behavior:
\begin{equation}
\frac{\partial_{\delta}N_u}{N_u}\approx-\frac{(N-1)\pi}{B[D/2-1,1/2]}\,\sin(\delta\pi)^{D-3}\;,
\end{equation}
with the beta function $B$. Fig.~\ref{fig:randomnetworks}B-C demonstrate the agreement between simulations and theory, which only depends on $N$ and $p$ (or $\langle k\rangle$) even for small values of $D$.

\paragraph{Number of clusters:}
We find that the number of clusters relative to the total number of nodes approaches a generic function for sufficiently large $D$, see Fig.~\ref{fig:randomnetworks}D-F. The number of clusters $N_c$ first rises due to cluster formation and later falls due to cluster merging as a function of $pN$. We can split the cluster number into two terms $N_c=N_{cf}-N_{cl}$, where $N_{cf}>0$ counts all events of cluster formation (\textit{i.e.}~two individual nodes become linked) and $N_{cl}>0$ counts all events of cluster merging. We find 
\begin{equation}
N_{cf}=N/4\,(1-(N_u/N)^2)\approx N/4\,(1-e^{-2pN})\,,
\end{equation}
with the number of unclustered nodes $N_u$ from \eqref{eq:NuApprox}.
The equation describes well the generic upward slope of $N_c$ as a function of the degree $\langle k\rangle=pN$ even for low dimensions (black dashed line).

In Erd\"os-R\'enyi random networks, there is a giant cluster, which arises at a percolation transition with $pN=1$. The giant cluster comprises a major fraction $S$ of all nodes given by the self-consistent equation \cite{newman2018networks}
\begin{equation}
S=1-e^{-\langle k\rangle S}\,.
\end{equation}
Using this, we find that the number of clusters peaks at $p_m$ with
\begin{equation}
N_{m}/N=(1-p_m N)\,((1-S)-e^{-p_m N})\,.
\end{equation}
As the maximum number $N_{m}>0$, it follows that $p_m N<1$.
Therefore, the maximum number of clusters is found below the percolation transition, where $S=0$ (Fig.~\ref{fig:randomnetworks}F). We can also identify a lower bound for $p_m$ as $3-4p_m N<e^{-p_m N}$ and thus $p_m N>0.5$.

\paragraph{Percolation transition:}
For a given value $\delta$ or $\langle k\rangle$, we might observe several clusters of various sizes $s_i$, which constitute a set $\mathbb{S}(\delta)=\{s_i\}$. We can rank the cluster sizes in decreasing order, such that $s_1=\max(\mathbb{S})$ and $s_2=\max(\mathbb{S}\setminus s_1)$. Fig.~\ref{fig:randomnetworks}G-H shows the behavior of the two largest cluster sizes $s_1/N$ and $s_2/N$ as a function of $\langle k\rangle=p(\delta,D)N$ for various dimensions $D$ and numbers of nodes $N$, respectively. Initially, all cluster sizes increase, yet eventually the system reaches a percolation transition, beyond only one giant cluster grows, while all other cluster sizes decrease. For a sufficiently large dimensions $D$ (yet, interestingly, independently of $N$), the percolation transition appears at $\langle k^*\rangle=1$. This is in analogy to random networks of the Erd\"{o}s-R\'{e}nyi type, for which a single connection per node on average is sufficient to generate a giant cluster \cite{bollobas2001random,newman2018networks}. In contrast, in lower dimensions $D$, the percolation transition of the random geometric network is shifted to higher values of the average degree.
Nevertheless, we still find a generic relationship $f(D)$ independent of $N$ to describe the percolation threshold (Fig.~\ref{fig:randomnetworks}I).
We determine the percolation threshold from the maxima of the cluster size curves for ranks $i>1$ (Fig.~\ref{fig:randomnetworks}L).

The critical point of a percolation transition is known for its universal properties, which allows to apply the identical theoretical framework to diverse systems. We observe a generic behavior of the cluster sizes even far away from the percolation transition. First, like in Erd\"{o}s-R\'{e}nyi random networks, the sizes of all clusters but the giant cluster scales logarithmically with $N$ (Fig.~\ref{fig:randomnetworks}J) \cite{newman2018networks}. Second, rescaling of the $p$ (or the mean degree $\langle k\rangle$) by its critical value yields a collapse of the curves for different dimensions onto a generic master curve (Fig.~\ref{fig:randomnetworks}K).

\paragraph{A conservative estimate on the cluster size:}
In real-world data sets like in RNA sequencing data, the data points corresponding to the nodes of a random geometric network might not be uniformly distributed across the entire hypersurface of dimension $D-2$ but correlated and constraint to a submanifold. Thereby, the same percolation threshold $\delta^*$ can be observed for very different manifolds. We identify two limiting cases, for which data points are constraint to either a hyperspherical cap or a lower dimensional hypersphere (gray region and gray line in the schematic of Fig.~\ref{fig:randomnetworks}M).  Fig.~\ref{fig:randomnetworks}M illustrates that the relationship between the mean degree and the distance $\delta$ derived in Eqs.~\ref{eq:connectionprob}-\ref{eq:meandegree} holds generically, when rescaling $\delta$ by the critical value $\delta^*$ of the percolation transition. Thus, for $\delta<\delta^*$, matching the dimension $D$ to a given $\delta^*$ results in larger values for $\langle k\rangle/\langle k^*\rangle$ in comparison to matching the cap size. A larger ratio $\langle k\rangle/\langle k^*\rangle$ yields larger cluster sizes according to Fig.~\ref{fig:randomnetworks}K.

In our clustering procedure, we attempt to devise a null model from a given value of $\delta^*$ and compare the expected largest cluster sizes with the observed cluster sizes.
Thus, adjusting the dimension $D$ to fit $\delta^*$ leads to a conservative estimate of the expected largest cluster sizes $s_1$. However, the estimates do not differ tremendously for the relevant regime close to the percolation transition and for typical parameter values, as exemplified in Fig.~\ref{fig:randomnetworks}N.

Furthermore, our clustering approach is based on constructing a local null model for the noise background based on a local percolation behavior. Yet, we do not know how many nodes constitute the local environment, although for typical data sets with weakly inhomogeneous backgrounds, the number is expected to be in the order of $\mathcal{O}(N/10)$. Again, we use a conservative estimate by constructing our null model using the total number of nodes $N$. This slightly overestimates the expected maximum cluster size $s_1$, yet only with a logarithmic correction (Fig.~\ref{fig:randomnetworks}J). The overestimation of $s_1$ by $N$ is partially counterbalanced by the effect of $D$ on $s_1$ discussed above (Fig.~\ref{fig:randomnetworks}M).

\paragraph{Clustering based on a local percolation criterion}
First, we compute the complete dendrogram of hierarchical single-linkage clustering. We assume that clusters of correlated genes merge with each other or with the noise background due to a local percolation behavior that sweeps through the system as a function of $\delta$. Thus, we construct a (conservative) local null model at each branching point $\delta$ by solving for $D$ in
\begin{equation} p(\delta,D)N=f(D)\,, \end{equation}
where $f(D)$ is the generic relationship shown in Fig.~\ref{fig:randomnetworks}I.
This allows us to compute the expected largest cluster size as a function of $\delta$ for a uniform distribution of $N$ nodes on a $D-1$-dimensional hypersphere.

Next, we check whether at least the branch corresponding to the smaller of the two merging clusters (or when both clusters are of equal size, the shorter branch) contains at least one cluster that violates the null model. This means there exists a size $s_1(\delta)$, for which $\delta<\delta_0(s_1)-\rho\,\sigma_0(s_1)$. Here, $\delta_0(s_1)$ and $\sigma_0(s_1)$ is the expected value for the distance and its standard deviation, at which the cluster size $s_1$ is reached in the null model. Note that $\delta(s_1)$ follows a normal distribution (Fig.~\ref{fig:randomnetworks}O). The parameter $\rho$ controls the false discovery rate of clusters with the standard choice $\rho=2$ or $\rho=3$. Clusters might be split in sub-clusters if the local percolation condition is fulfilled for lower values of $\delta$.

An identified cluster is tracked until it reaches the the expected largest cluster size of the null model. This closure of the cluster avoids to accumulate too many false cluster members from the background. As a side note, one obvious extension to refine the clustering process would be to add an additional parameter that controls the accumulation of false cluster members.

\paragraph{Generating synthetic data:}
To generate uniformly distributed data on the hypersurface of a $(D-1)$-dimensional hypersphere, each data point $\vec{v}_i$ is computed as
normalized vector of normally distributed random variables $g$ with a zero mean: $\vec{v}_i=\vec{g}_i/\sqrt{\sum_j g_{ij}^2}$, where the vector $\vec{g}_i=[g_{i1},...,g_{iD}]$, corresponding to $D$ independent measurements of a gene. This is equivalent to the z-score divided by $\sqrt{D}$: $\vec{v}_i= (\vec{g}_i-\langle\vec{g}_i\rangle)/\sqrt{\langle\vec{g}_i^2\rangle-\langle\vec{g}_i\rangle^2}/\sqrt{D}$. The brackets $\langle\rangle$ denote the average across measurements.

Clusters of correlated data points can be generated by adding a $D$-dimensional vector to all the respective random vectors before z-scoring. For example, the clusters of data points in Fig.~\ref{fig:method}D are obtained by generating $N=1500$ vectors $\vec{g}_i=[g_{i1},g_{i2},...,g_{i10}]$ as above and then replacing $\vec{g}_i\rightarrow\vec{g}_i-[0,4/9,8/9,...4]$ for $1\leq i\leq 50$ before computing the z-score. The upper dendrogram of Fig.~\ref{fig:method}F is obtained by from $N=1000$ vectors $\vec{g}_i=[g_{i1},g_{i2},g_{i3},g_{i4}]$ and then replacing $\vec{g}_i\rightarrow\vec{g}_i+[8,8,-8,-8]$ for $1\leq i\leq 25$ and $\vec{g}_i\rightarrow\vec{g}_i+[-8,-8,8,8]$ for $26\leq i\leq 50$.

\begin{figure*}[t!]
\includegraphics[width=1\linewidth]{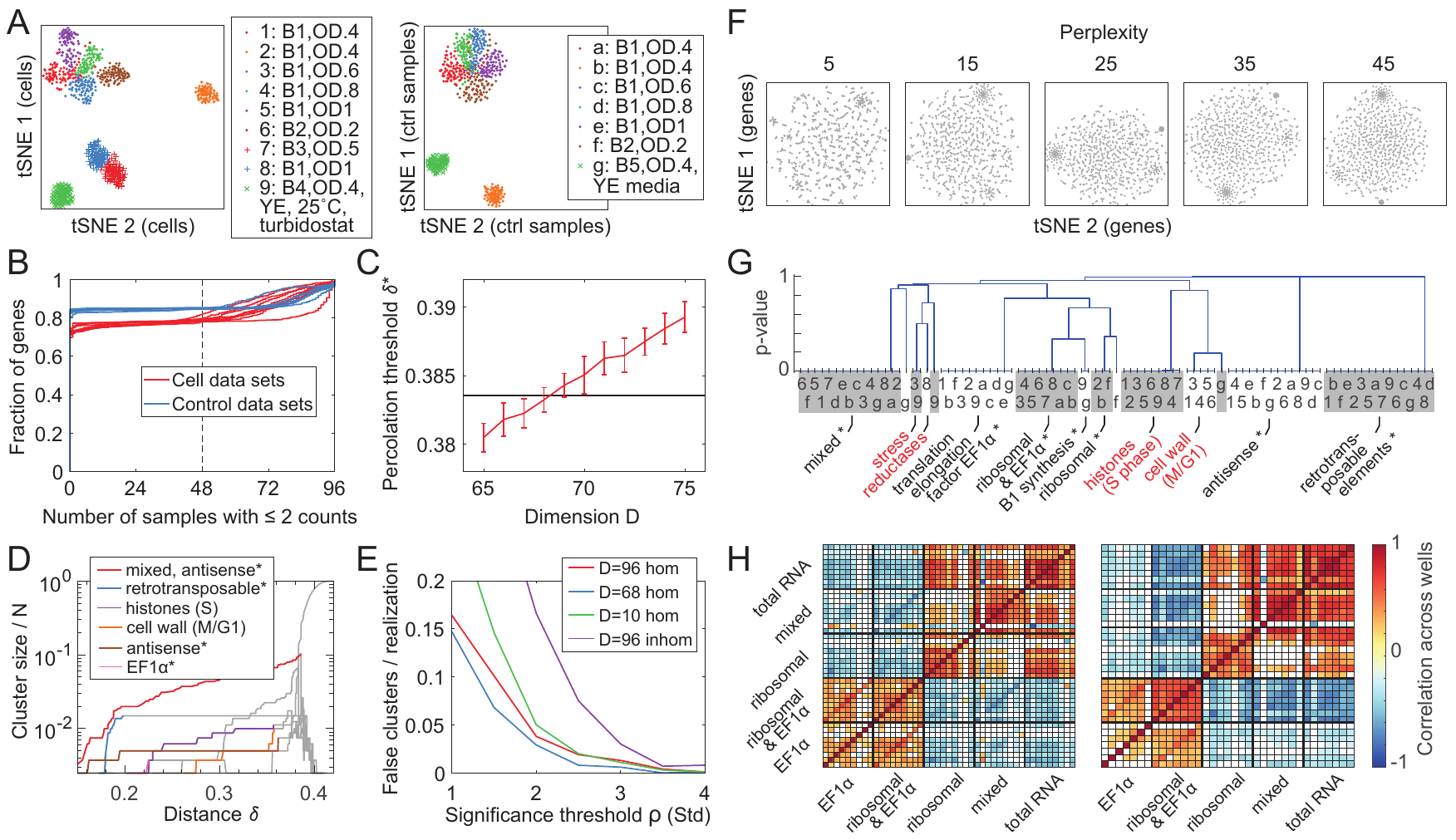}
\caption{{\bf Clustering single-cell variation in fission yeast across data sets.} (A) t-SNE plot of individual samples (dots, crosses) across all data sets for either single cell and control bulk data. Yeast cells are cultured in a flask at $32^{\circ}$C in EMM2 media (unless otherwise specified for batches B4 and B5). Cells were collected at the specified densities from OD 0.2 to OD 1.0.
(B) Fraction of genes with a certain number of low counts shows a plateau, resulting in a natural cut-off criterion (dashed line).
(C) Uncorrelated data with D=68 possess a percolation threshold at approximately the same value as the yeast data set 1  (black line).
(D) Growth of the identified clusters as a function of $\delta$ for data set 1. Gray lines are the seven largest cluster sizes $s_i/N$ as in Fig.~\ref{fig:yeast}C.
(E) False positive clusters per realization of random data without sets of correlated genes as a function of the clustering parameter $\rho$ for homogeneous distributions in the relevant dimensions and for a globally inhomogeneous distribution approximately matching the inhomogeneity of the noise background.
(F) t-SNE plots of the genes in data set 1 look similarly noisy independent of the perplexities (shown from 5 to 45).
(G) Meta-clustering of all gene clusters from all data sets (specified by respective number or letter) with gene annotation (asterisk denotes overlap with clusters from control data sets, red cluster annotations are not found in the control). p-values obtained from a Fisher test.
(H) Pearson correlation across wells between all single cell and control data sets, respectively, for the mean z-score of all gene clusters and the total RNA count shows that obtained counts are high or low depending on the sequenced well.
}\label{fig:yeastall}
\end{figure*}

To generate an inhomogeneous background as shown in the lower dendrogam of Fig.~\ref{fig:method}F, we add a slightly different vector to each normally distributed gene vector before z-scoring. Specifically in Fig.~\ref{fig:method}F, we used a set of $N=1000$ vectors of the from $\vec{g}_i=\vec{g}_i+a_i[-1,-1/3,1/3,1]$ with $N$ equally spaced values $a_i$ ranging from $0$ to $1$. For the inhomogeneous distribution in Fig.~\ref{fig:yeastall}E (purple line), we use $\vec{g}_i=\vec{g}_i+a_i[-0.5,...,0.5]$.

$N$ data points in a hyperspherical cap covering (approximately) a fraction $1/\xi$ of the hypersurface are obtained by initially generating $N\xi$ data points across the entire hypersphere, yet subsequently only selecting the $N$ points closest to the first generated data point.

\paragraph{On the clustering of fission yeast:}
We analyze RNA sequencing measurements from fission yeast, which consists of 9 individual data sets of fission yeast cells under six different conditions and additionally 7 control data sets of bulk samples under matching conditions. The t-SNE plot of the individual cells/samples in Fig.~\ref{fig:yeastall}A illustrates that all data sets (even data sets under identical conditions) cluster separately and similar or identical conditions are not necessarily close in t-SNE space. Thus, we analyze each data set independently, focusing on data set 1.

About $80\%$ of the genes have more then 2 RNA counts in most of the 96 samples, the remaining genes have lower counts in almost all samples Fig.~\ref{fig:yeastall}B). Thus, we only consider genes with more than 2 counts in at least half of the samples (dashed line) as a natural cut-off.

The data sets are very noisy as exemplified by a t-SNE projection for data set 1, independent of the perplexity values (Fig.~\ref{fig:yeastall}F). We determine the effective dimension $D=68$ of uncorrelated data that matches the percolation transition of data set 1 (solid black) (Fig.~\ref{fig:yeastall}C). This allows us to compare the observed cluster sizes in data set 1 with the expected maximum cluster size in uncorrelated data (Fig.~\ref{fig:yeast}C). The actual growth of the later identified clusters are shown in Fig.~\ref{fig:yeastall}D.

For the clustering, we use a significance threshold of $\rho=3$, which yields $6$ clusters with distinct annotations. If we choose $\rho=2$, we obtain the identical clusters plus an additional one, for which, however, we cannot find coherent gene annotations. In fact, we can estimate the false discovery rate for the case of random data without clusters. Homogeneous (uniform) distributions of points on the hypersphere in the range of $D$ covered by the clustering as well as an inhomogeneous distribution with a similar dendrogram structure as the yeast data shows that the expected number of falsely discovered clusters per experiment is below $0.05$ for $\rho=3$ (Fig.~\ref{fig:yeastall}E).

Finally, we compare the clustering result for each individual data set by a meta-clustering based on the overlap between clusters of all data sets (Fig.~\ref{fig:yeastall}G). We use hierarchical average-linkage clustering with the p-values from a Fisher test as the distance measure and a cut-off of $0.01$. This shows that we have repeatedly identified similar gene clusters in different data sets with a large mutual overlap and defining characteristics. We merge the gene clusters from different data sets by only taking genes into account, which are found in more than one data set.

The fact that many of the clusters are also found in control data sets poses the question whether they might originate to some extent from measurement artifacts. For example, the measured counts of four of the clusters (yet not the remaining clusters) correlate across wells of the 96-well plate between data sets (Fig.~\ref{fig:yeastall}H). The dependence on the well seems further be related to the differences in sequencing depth between wells. Nevertheless, these clusters still have concise functional annotation and might be interesting to explore further.

\begin{figure}[t!]
\includegraphics[width=0.99\linewidth]{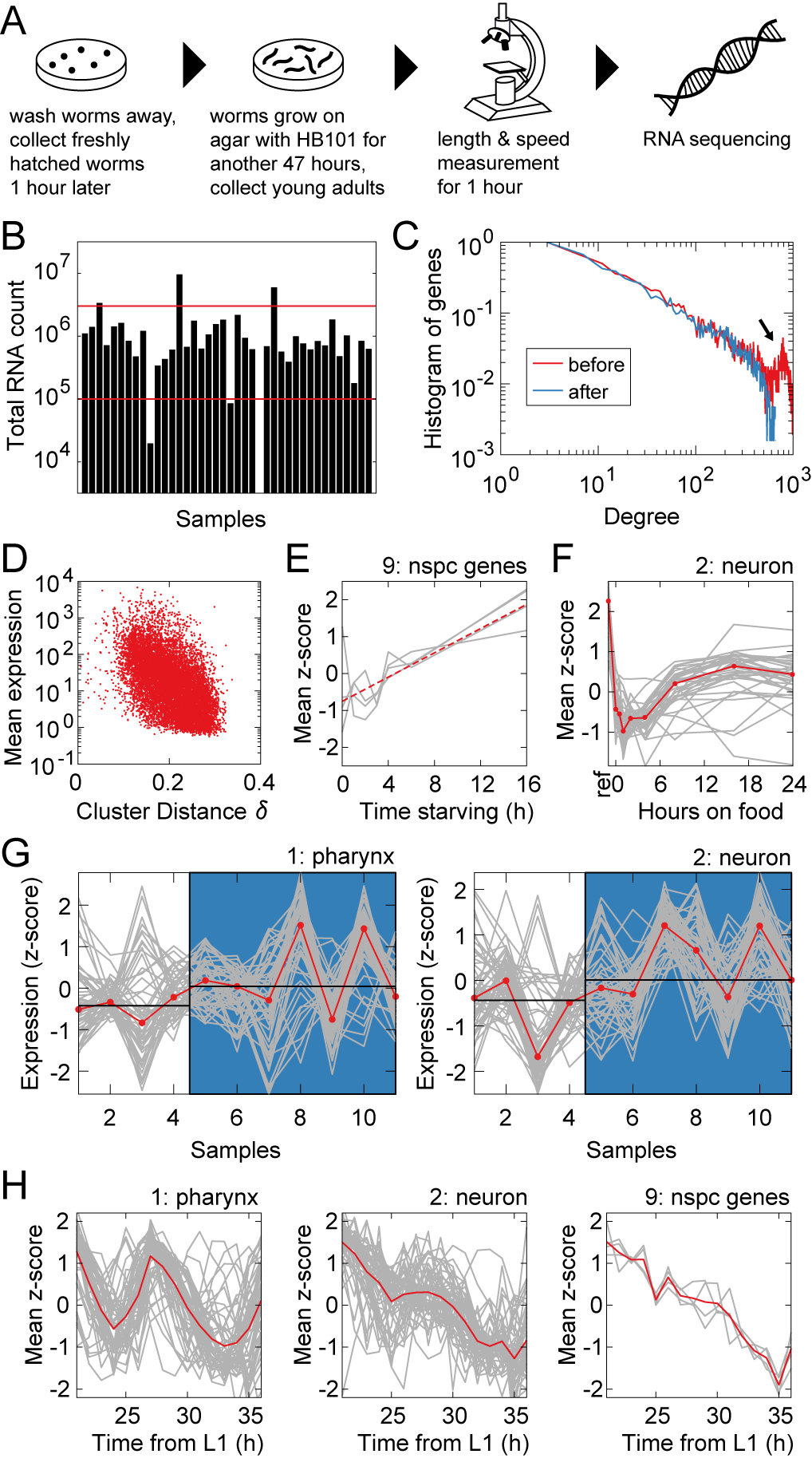}
\caption{{\bf Extracting and interpreting clusters from whole-worm RNA sequencing data of {\it C. elegans}}. (A) Experimental protocol. (B) We exclude samples with very high and very low RNA counts (red lines). (C) Removing samples with high sequencing depth removes hubs of genes, which correlate with a large fraction of other genes (exemplified by a degree distribution for $\delta=0.25$).
(D) Lowly expressed genes contribute to clusters higher up in the clustering dendrogram.
(E) Genes of cluster 9 are upregulated during starvation, analysis of data from ref.~\cite{harvald2017multi}.
(F) Transient expression dynamics of the neuron genes (cluster 2) in worms on food that have previously been subject to three rounds of 1h long starvation periods (separated by 30min feeding periods), fold change computed relative to reference worms without a starvation experience (first data point), our analysis of data from ref.~\cite{freytag2017genome}
(G) Comparison of pharynx and neuron genes between well-fed L1 worms and L1 worms with a starvation experience (dauer) three generations earlier, our analysis of data from ref.~\cite{webster2018transgenerational}. Wilcoxon rank-sum test yields a p-value of 0.07 (pharynx) and 0.04 (neurons).
(H) The neuron, the pharynx and the nspc cluster show distinct dynamics during development, although in our measurements the development is not the major driver of expression variability.}\label{fig:celeganspre}
\end{figure}

\paragraph{On the clustering of {\it C. elegans} data:}
We produce a RNA sequencing data set using 40 young adult {\it C. elegans} worms, for which we also measure the size and the mean speed of the worms (Fig.~\ref{fig:celeganspre}A). We exclude 3 samples with very low sequencing depth and 3 samples with very high high sequencing depth, yielding $D=34$ samples for the subsequent analysis (Fig.~\ref{fig:celeganspre}B). The samples with high sequencing depth are excluded because they appear to dominate the data set by generating global correlations, as shown in \ref{fig:celeganspre}C. The degree of each gene is the number of other genes within a given distance (\textit{i.e.}~correlation) threshold. Thus, the degree distribution in Fig.~\ref{fig:celeganspre}C shows that before excluding the samples with high RNA counts (red) there are many genes which correlate with a large fraction of the genes in the data set (arrow). After removing these samples, the degree distribution follows approximately a power law with an exponential cut-off, indicating the presence of small groups of correlated genes. Completely, uncorrelated data (\textit{i.e.}~uniformly distributed on the hypersphere) would yield a binomial distribution.
Furthermore, we remove all genes with zero counts in at least half of the samples, which results in $N=11629$ genes. This filtering step is not essential (most gene clusters are found irrespectively), yet it increases the sensitivity of the clustering analysis.

The background noise is clearly inhomogeneous with a smeared out percolation behavior, which is indicated by the tilt of the dendrogram in Fig.~\ref{fig:celegans}A in comparison to uncorrelated data drawn from a normal distribution. While highly expressed genes contribute to the early percolation behavior, lowly expressed genes are more uniformly distributed (\textit{i.e.}~more uncorrelated) and are clustered only at large $\delta$ values (Fig.~\ref{fig:celeganspre}D). For the clustering, we choose $\rho=2$($\sigma$) to control for false discovery of clusters. This slightly less conservative choice in comparison to the yeast data set still yields a set of clusters all of which are biologically interpretable.

\begin{figure}[tb]
\includegraphics[width=1\linewidth]{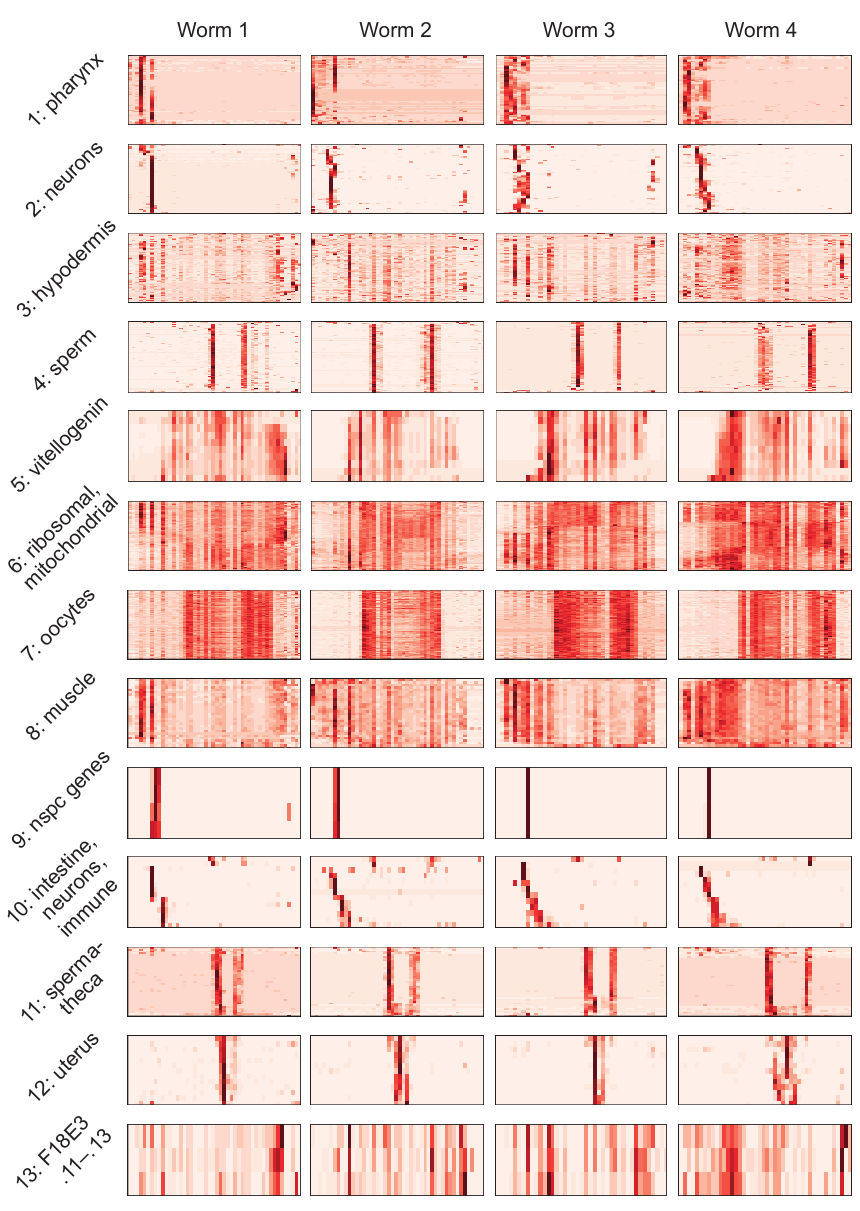}
\caption{{\bf {\it C.~elegans} clusters and their respective anatomical localization.} Spatially resolved tomoseq data from ref.~\cite{ebbing2018spatial} of 4 young adult worms (red: high gene expression). Gene sorted for each data set independently to optimize similarity between neighbors.}\label{fig:tomoseq}
\end{figure}

Fig.~\ref{fig:tomoseq} shows the spatial patterns of all the clusters along four different worms. The results are highly consistent across individuals and coherent with the functional annotations. A clear annotation is only missing for cluster 13. This cluster consists of 3 uncharacterized genes, which are direct neighbours on the chromosome. Interestingly, this is the only clusters that shows a significant correlation with the mean speed of the worm.

Cluster 1 (pharynx), cluster 2 (neurons) and cluster 9 (nspc genes) display an overall very similar expression variation in our measurement as well as very similar dynamics during starvation (Fig.~\ref{fig:celegans}C and D, Fig.~\ref{fig:celeganspre}E). 
Could a starvation experience be the underlying cause for the common expression variation of these three clusters? Worms did not have access to food during the length measurement before sequencing, which lasted for 52 to 67 minutes. The small differences in starvation time between samples do not correlate with the expression variations. Thus, an immediate response to feeding conditions might be unlikely, however we find evidence that the clusters might emerge from long-lasting memory effects.
First, the neuron cluster possesses a transient expression dynamics over several hours, when being transferred from starvation to feeding conditions (Fig.~\ref{fig:celeganspre}F). Second, the feeding memory might even extend across several generation. Our pharynx and neuron cluster show very similar expression changes in an experiment by Webster {\it et al.}, which compares L1 worms with and without a harsh starvation experience three generations ago (Fig.~\ref{fig:celeganspre}G). However, additional experiments are needed to confirm this possible trans-generational effect in the future.
Third, the observations are complemented by the fact that one member of the neuron cluster, daf-7, has been described as a key player in behavioral feeding response as well as acquired (trans-generational) pathogenic avoidance behavior \cite{hilbert2017sexually, moore2019piwi}.
While the pharynx, the neuron and the nspc gene clusters show many similarities in their expression behavior, they are identified as individual clusters with a distinct spatial pattern. This might require other sources of variability such as different temporal dynamics during development (Fig.~\ref{fig:celeganspre}H). 

%
%
%
%
%
%
%
%
%

\matmethods{

\subsection*{Analysis of the yeast data set}
A previously published data set of RNA sequencing measurements in {\it Schizosaccharomyces pombe} was obtained from ref.~\cite{saint2019single}. We only use the growth conditions (9 single cell data sets and 7 bulk control data sets). Each data set contains 96 samples. We remove duplicated entries in the table (sample 67/68 in data set 2 and sample 76/77 in data set 9). Finally, we filter out all genes with less then 3 counts in more than half of the samples, which reduces the number of genes to approximately $75\%$ in single cell data sets and $85\%$ in control data sets (Fig.~\ref{fig:yeastall}B).  Finally, we apply our clustering procedure using $\rho=3$($\sigma$) to control for the false discovery rate of clusters. To compare clustering results from individual data sets, we apply a meta-clustering based on the overlap between clusters. We define a similarity score based on the Fisher test, which provides a p-value for the null hypothesis to obtain the overlap by chance.

\subsection*{RNA sequencing of {\it C. elegans} worms}
{\it C. elegans} worms of the Bristol N2 strain were grown on NGM agar plates coated with E. coli HB101 at $20^{\circ}$C under standard conditions \cite{lewis1995basic}. Worms were synchronized 48h prior to the experiment (Fig.~\ref{fig:celeganspre}A). Culture plates were washed with M9 medium to remove all hatched worms, leaving eggs behind. L1 larvae were collected 0-1h after hatching and seeded onto fresh NGM plates where they were grown for 47-48 additional hours. Approximately, 1h before the sequencing, worms were put on fresh NGM plates without food and size and worm motility has been measured using a FLIR GS3-U3-123S6M-C camera with a resolution of 10.8um/pixel.
After the recording, the worms were treated with Trizol, flash freezed with liquid nitrogen and stored at -$80^{\circ}$C.
mRNA extraction, barcoding, reverse transcription, and in vitro transcription were performed according to the CEL-seq protocol \cite{hashimshony2012cel}  using Message Amp II kit (Ambion). Illumina sequencing libraries were subsequently prepared according to the CEL-seq2 protocol \cite{hashimshony2012cel} using the SuperScript® II Double-Stranded cDNA Synthesis Kit (Thermofisher), Agencourt AMPure XP beads (Beckman Coulter), and randomhexRT for converting aRNA to cDNA using random priming. The libraries were sequenced paired-end at 75 bp read length on an Illumina HiSeq 2500.
For each read, the worm-specific barcode and unique molecular identified are detected in the first 8 nucleotides and 4 nucleotides, respectively, of the first read. Second reads were aligned to the C. elegans reference transcriptome, which was compiled from the C. elegans reference genome WS249 \cite{ebbing2018spatial}. Only reads that uniquely mapped to the transcriptome and that have a proper worm-specific barcode are used for downstream analysis.
A custom wrapper (MapAndGo2) was used for the alignment around BWA MEM \cite{ebbing2018spatial, li2010fast}. Raw data were processed, removing amplification duplicates \cite{grun2014validation} and analyzed using MATLAB. 34 samples containing at least $10^5$ and less than $3\cdot 10^6$ total transcripts were used for follow-up analysis (see Fig.~\ref{fig:celeganspre}B-C).
We remove genes with zero counts in at least half of the samples. Our clustering procedure uses a threshold $\rho=2$($\sigma$) to control for the false discovery of clusters.
}

\showmatmethods{} 

\acknow{
We thank the members of the Shimizu group and the Stephens group and all members of the NWO/FOM NOISE consortium for stimulating discussions. We acknowledge funding by the Netherlands Organisation for Scientific Research (Nederlandse Organisatie voor Wetenschappelijk Onderzoek; NWO) via NWO/FOM Program Grant no. 161. TSS would like to thank the MRC Laboratory for Molecular Biology, Cambridge, and the MRC London Institute for Medical Sciences for their hospitality during several months in which part of this work was carried out.}

\showacknow{} 

\bibliography{ClusteringPaper}

\end{document}